\documentclass[%
 reprint,
 amsmath,amssymb,
 aps,
]{revtex4-1}

\usepackage{graphicx}
\usepackage{dcolumn}
\usepackage{bm}
  \usepackage{lineno}

\begin{document}

\preprint{APS/123-QED}

\title{J/$\psi$ production as a function of event multiplicity in pp collisions at $\sqrt{s}$ = 13 TeV using EMCal-triggered events with ALICE at the LHC}

\author{Cristiane Jahnke for the ALICE Collaboration}
\affiliation{%
 Physics Department, University of S\~ao Paulo
}%


\date{\today}


\begin{abstract}
The study of the J/$\psi$ production in pp collisions provides important information on perturbative and non-perturbative quantum chromodynamics. Using high multiplicity pp events, we can study how charmonium production depends on the event activity. 
These measurements are used to investigate the possible influence of multiple partonic interactions to the J/$\psi$ production and the interplay between soft and hard processes.
 In this work we report on studies of J/$\psi$ production as a function of event multiplicity in pp collisions  at  $\sqrt{s}$ = 13 TeV at mid-rapidity with ALICE. 
The J/$\psi$ are reconstructed via their dielectron decay channel in events where at least one of the decay electrons was triggered on by the Electromagnetic Calorimeter (EMCal). 
The availability of a high-$p_{\rm T}$ electron trigger enhances the sampled luminosity significantly relative to the available minimum-bias triggered data set and extends the $p_{\rm T}$ reach for the J/$\psi$ measurement. Using these data, the J/$\psi$ measurement is performed in the transverse momentum interval $8 < p_{\rm T} < 30~\textrm{GeV}/c$.

\begin{description}
\item[PACS numbers]
25.75.Dw, 25.75.Cj, 14.65.Dw, 13.20.Fc

\end{description}

\end{abstract}

\pacs{Valid PACS appear here}
\maketitle


\section{\label{sec:level1} Introduction}

The production of charmonium (bound state of c and $\bar{\rm c}$ quarks) at RHIC and LHC energies is not yet fully understood. The production of the heavy-quark pair can be described by perturbative Quantum ChromoDynamics (pQCD) while its hadronisation into a quarkonium state is a non-perturbative process \cite{Fritzsch:1977ay, Baier:1981uk, Andronic:2015wma}. 
The study of their production as a function of charged-particle multiplicity is an interesting observable since it can give us insights into parton level and into the interplay between soft and hard processes.
 In particular, the Multiple Parton Interaction (MPI) \cite{Sjostrand:1987su}, which was in general used to describe soft processes, is currently also used to describe hard processes such as quarkonium production. 
Thus, this study can shed light on possible connections between the hard production of heavy quarks and soft particle production \cite{Porteboeuf:2010dw}.
 Additionally, pp measurements provide a baseline for proton--nucleus and nucleus--nucleus collisions allowing the quark--gluon plasma properties to be studied.
 In this work we study the J/$\psi$ production as a function of charged-particle multiplicity in pp collisions at a centre-of-mass energy of $\sqrt{s}$ = 13 TeV at mid-rapidity with ALICE \cite{Aamodt:2008zz}.

\section{\label{sec:level1}Experimental setup and analysis strategy}

The data set used in this analysis was recorded by ALICE using three different triggers: a minimum-bias (MB) trigger, a high multiplicity (HM) trigger and an Electromagnetic Calorimeter (EMCal) trigger \cite{Abeysekara:2010ze}.
The MB trigger used the V0 detector, which consists of two arrays of scintillator tiles at both forward, 2.8 $< \eta <$ 5.1 (V0A) and backward, $-$3.7 $< \eta <$ $-$1.7 (V0C) pseudorapidity regions.
It required a coincidence of signals in the V0A and V0C detectors.
The HM trigger is based on a large deposited charge in the  V0 detector. 
The EMCal trigger is used to select events containing high-$p_{\rm T}$ electrons and photons.
The two innermost layers of the Inner Tracking System (ITS) were used to estimate the charged-particle multiplicity of the event. These layers comprise the Silicon Pixel Detector (SPD) at radii
3.9 cm and 7.6 cm from the nominal interaction point, located at mid-rapidity surrounding the beam pipe. In this analysis, only events where at least one track was reconstructed by the SPD were used (INEL$>$0).

\begin{figure}[h!]
\includegraphics[width=0.5\textwidth]{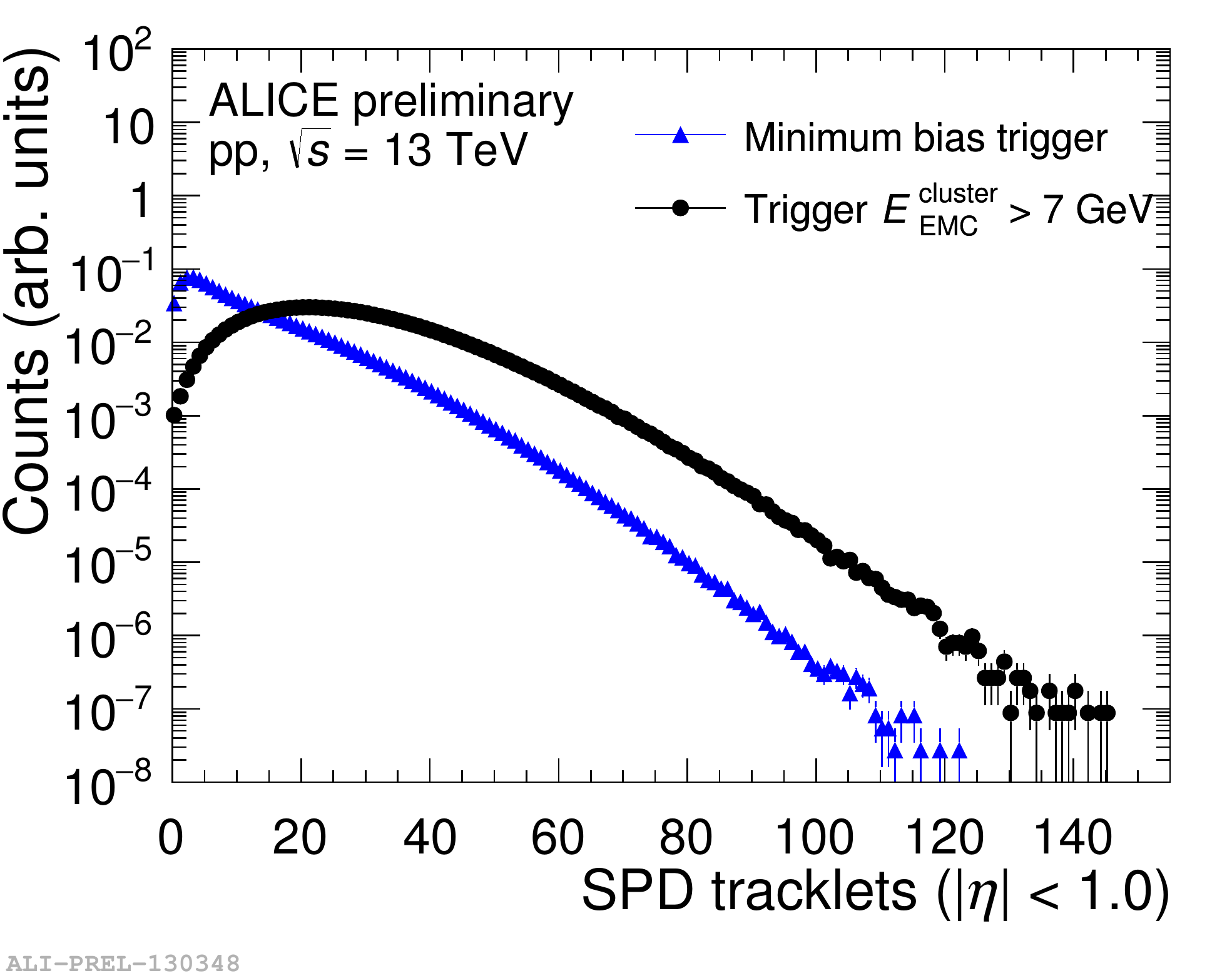}
\caption{\label{fig:mult} Multiplicity distribution based on SPD tracklets for minimum-bias triggered events and EMCal-triggered events in pp collisions at $\sqrt{s}$ = 13 TeV.}
\end{figure}

Figure \ref{fig:mult} shows the number of SPD tracklets in a MB data set compared with the EMCal-triggered data set.

\begin{figure}[h!]
\includegraphics[width=0.5\textwidth]{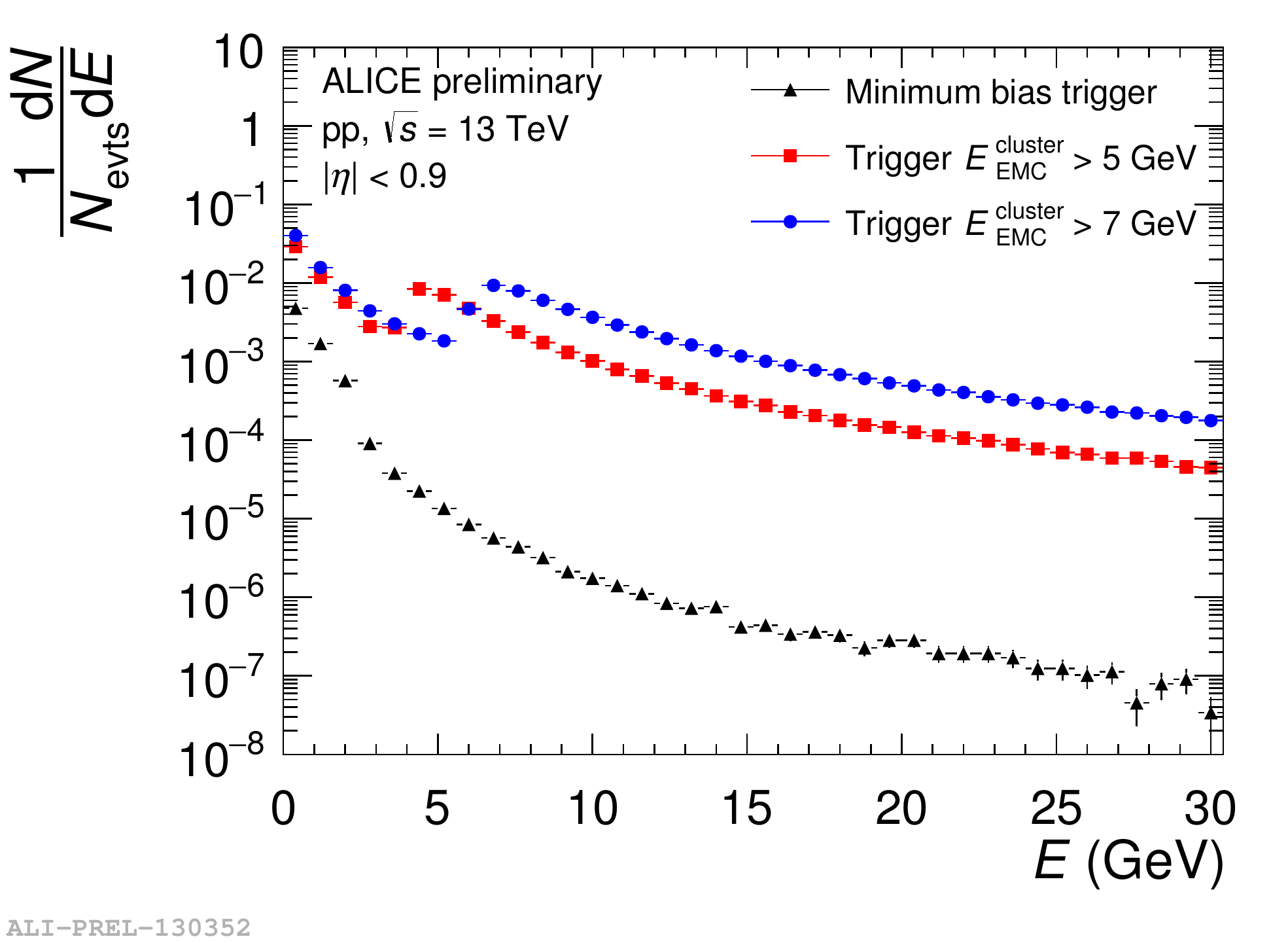}
\caption{\label{fig:cls} Cluster energy distribution for EMCal-triggered data ($E_{\rm EMC}^{\rm cluster} > 7$ GeV and $E_{\rm EMC}^{\rm cluster} > 5$ GeV) and for MB-triggered data in pp collisions at $\sqrt{s}$ = 13 TeV.}
\end{figure}

The electron and positron candidates used to reconstruct J/$\psi$ were identified by the Time Projection Chamber (TPC) and their trajectories were reconstructed using the combination of tracks on TPC and ITS detectors.  These tracks are matched with EMCal cluster, which measures the energy ($E$) of the particles and the electrons/positrons are identified based on $E/p$ where $p$ is the momentum of the tracks. Since electrons deposit all their energy in the EMCal and have a negligible mass, electrons are expected to have an $E$/$p$ value around one.
The J/$\psi$ are reconstructed in events where at least one of the decay electrons fired the EMCal trigger. The electrons/positrons are identified using the TPC and we require at least one of the legs of the J/$\psi$  in the EMCal, with a cluster energy above the trigger threshold and a ratio of energy over momentum in  $0.8 < E/p < 1.3$. 
 The availability of a high-$p_{\rm T}$ electron trigger enhances  significantly the sampled luminosity relative to the available sample of minimum-bias (MB) triggered data, extending the reached $p_{\rm T}$ for the J/$\psi$ measurement. Fig. \ref{fig:cls} shows the cluster energy distribution for two different trigger thresholds (at $E_{\rm EMC}^{\rm cluster} > 5$ GeV and at $E_{\rm EMC}^{\rm cluster} > 7$ GeV) compared with a MB triggered data set.
 
\begin{figure}[h!]
\includegraphics[width=0.5\textwidth]{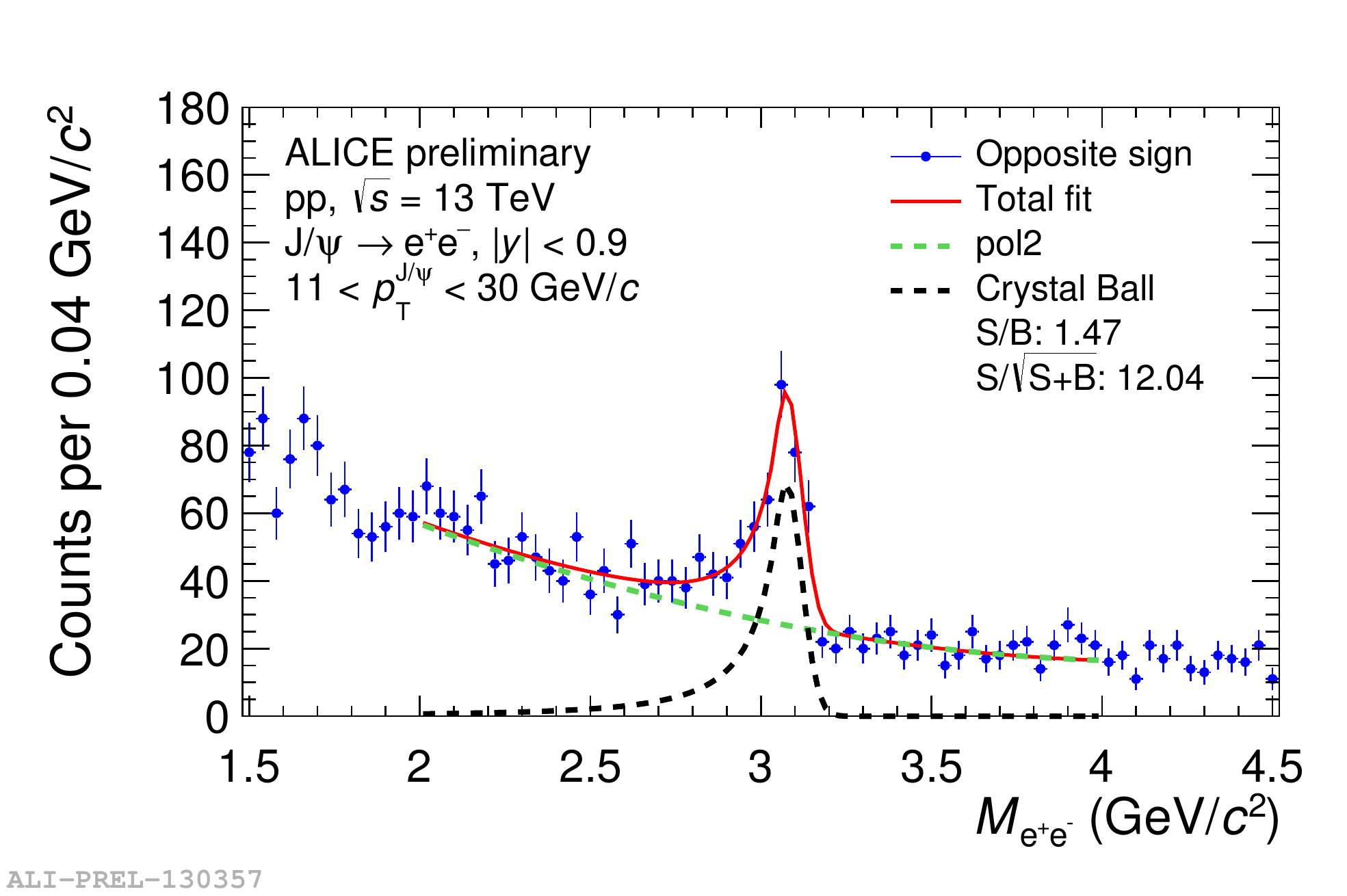}
\caption{\label{fig:invmass} Invariant mass spectrum of J/$\psi$ candidates in pp collisions at $\sqrt{s}$ = 13 TeV.}
\end{figure}

Figure \ref{fig:invmass} shows as an example J/$\psi$ candidates in the transverse momentum interval $11 < p_{\rm T} < 30~\textrm{GeV}/c$. 
The J/$\psi$ yield is obtained after subtracting the combinatorial background, which is described by a second-order polynomial. The background fit is performed excluding the peak region $ 2.5 < m_{\rm e^{+}e^{-}} < 3.3 $ GeV/$c^2$ and the  J/$\psi$ yield is calculated in the mass range  $ 2.92  < m_{\rm e^{+}e^{-}} < 3.16$ GeV/$c^2$. 

The signal was extracted in five different multiplicity bins $i$ as shown in Eq. \ref{eq:yield}. 

To obtain the J/$\psi$ self-normalized yield ($\frac{\rm d \it N_{\rm J/\psi}/\rm d \it y}{\langle \rm d \it N_{\rm J/\psi}/\rm d \it y\rangle_{\rm INEL>0}}$), the signal in EMCal-triggered data per event, in each multiplicity bin ($N_{\rm J/\psi}^{i}/N_{\rm evts}^{i}$), is normalized to the observed signal per event integrated over multiplicity ($\langle N_{J/\psi}\rangle^{\rm integrated}/N_{\rm evts}^{\rm integrated}$):

\begin{equation}
\frac{\rm d \it N_{\rm J/\psi}/\rm d \it y}{\langle \rm d \it N_{\rm J/\psi}/\rm d \it y\rangle_{\rm INEL>0}}= \frac{ \it N_{\rm J /\psi}^{i}/\it N_{\rm evts}^{i}}{\langle \it N_{\rm J /\psi}\rangle^{\rm integrated}/\it N_{\rm evts}^{\rm integrated}} .
\label{eq:yield}
\end{equation}

The number of events in each multiplicity bin is corrected by the rejection factor of the EMCal trigger and the J/$\psi$ yields are corrected for the vertex finding efficiency (97.5\%).

The self-normalised charged-particle multiplicity ($\frac{\rm d \it N_{\rm ch}/\rm d\eta}{\langle \rm d \it N_{\rm ch}/\rm d\eta \rangle_{\rm INEL>0}}$) is defined as the average number of charged particles per unit rapidity in a multiplicity bin $i$ divided by the number of charged particles per unit rapidity in inelastic events with at least one charged particle at mid-rapidity ($\rm INEL>0$):


\begin{equation}
\frac{\rm d \it N_{\rm ch}/\rm d\eta}{\langle \rm d \it N_{\rm ch}/\rm d\eta \rangle_{\rm INEL>0}}= \frac{  \it N_{\rm ch}^{i} }{\langle \it N_{\rm ch}\rangle^{\rm MB}}.
\label{eq:ch}
\end{equation}

\section{\label{sec:level1}Results}

\begin{figure}[h!]
\includegraphics[width=0.5\textwidth]{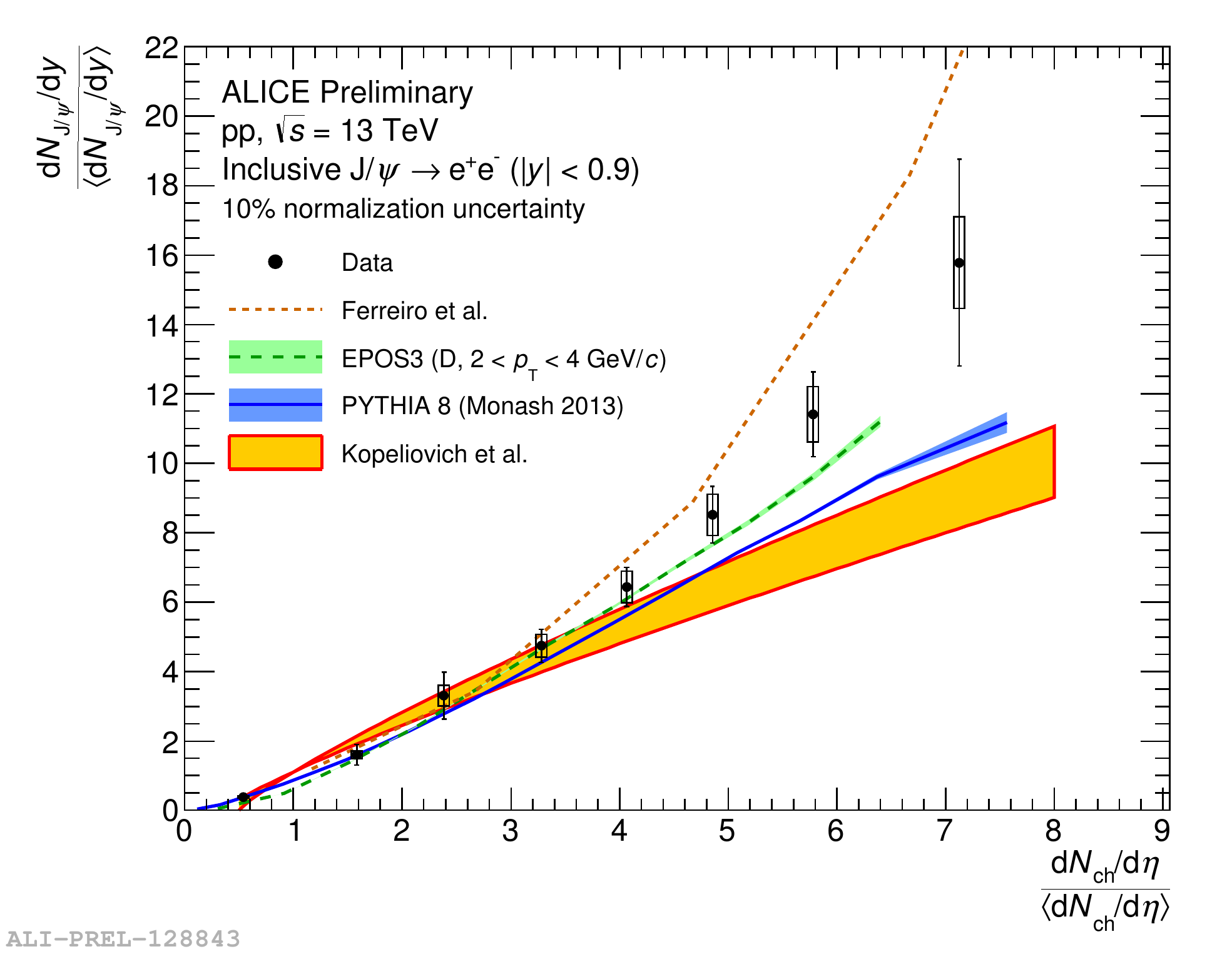}
\caption{\label{fig:result1} Multiplicity dependence of inclusive J/$\psi$ production at mid-rapidity in pp collisions at $\sqrt{s}$ = 13 TeV. The results are compared with several theoretical models: PYTHIA 8 \cite{Sjostrand:2007gs}, EPOS 3 \cite{Drescher:2000ha}, Ferreiro et al. \cite{Ferreiro:2012fb} and Kopeliovich et al. \cite{Kopeliovich:2013yfa}.}
\end{figure}

Figure \ref{fig:result1} shows the self-normalized inclusive J/$\psi$ yield as a function of the self-normalized charged-particle multiplicity in pp collisions at $\sqrt{s}$ = 13 TeV. 
These results were obtained using the ALICE high-multiplicty trigger and are integrated over $p_{\rm T}$ \cite{Weber:2017hhm}. 
An increase of J/$\psi$ yield stronger than linear with respect to the charged-particle multiplicity is observed. This increase of the yields of J/$\psi$  and D-meson were also observed in pp collisions at $\sqrt{s}$ = 7 TeV \cite{Abelev:2012rz, Adam:2015ota}.

The results are compared with several theoretical models: PYTHIA 8 \cite{Sjostrand:2007gs}, EPOS 3 \cite{Drescher:2000ha}, Ferreiro et al. \cite{Ferreiro:2012fb} and Kopeliovich et al. \cite{Kopeliovich:2013yfa}. 
Qualitatively, all models can reproduce the stronger than linear increase of the J/$\psi$ yield with respect to the event multiplicity. 
However, quantitatively, the model that best describes the data is  EPOS 3 \cite{Drescher:2000ha}, which is a model that includes a hydrodynamical expansion of the system and MPI. 
EPOS 3 uses QCD-inspired field theory for calculating cross sections and particle production. It also uses a unified treatment of soft and hard scattering: no fundamental cutoff parameter is used to define a border between soft and hard scattering \cite{Drescher:2000ha}. 

The model from Ferreiro et al.  slightly overestimates the increase observed in data at the highest multiplicities. This model assumes saturation of soft particle production and string interactions (percolation model). It assumes that all projectiles have a finite spatial extension and collides at finite impact parameter by means of elementary parton-parton collisions \cite{Ferreiro:2012fb}. Kopeliovich et al. \cite{Kopeliovich:2013yfa} assumes that hadron multiplicities larger than the mean value in pp collisions can be reached using higher Fock states in the proton, which contains an increased number of gluons. This model underestimates the results. PYTHIA 8 also underestimates the results at high multiplicities. It uses MPI and saturation of soft particle production via color reconnection. The color reconnection is used in the final state, in which there is a certain probability for the partons of two sub-scatterings to have their colors interarranged in a way that reduces the total string length \cite{Sjostrand:2007gs}.

\begin{figure}[h!]
\includegraphics[width=0.5\textwidth]{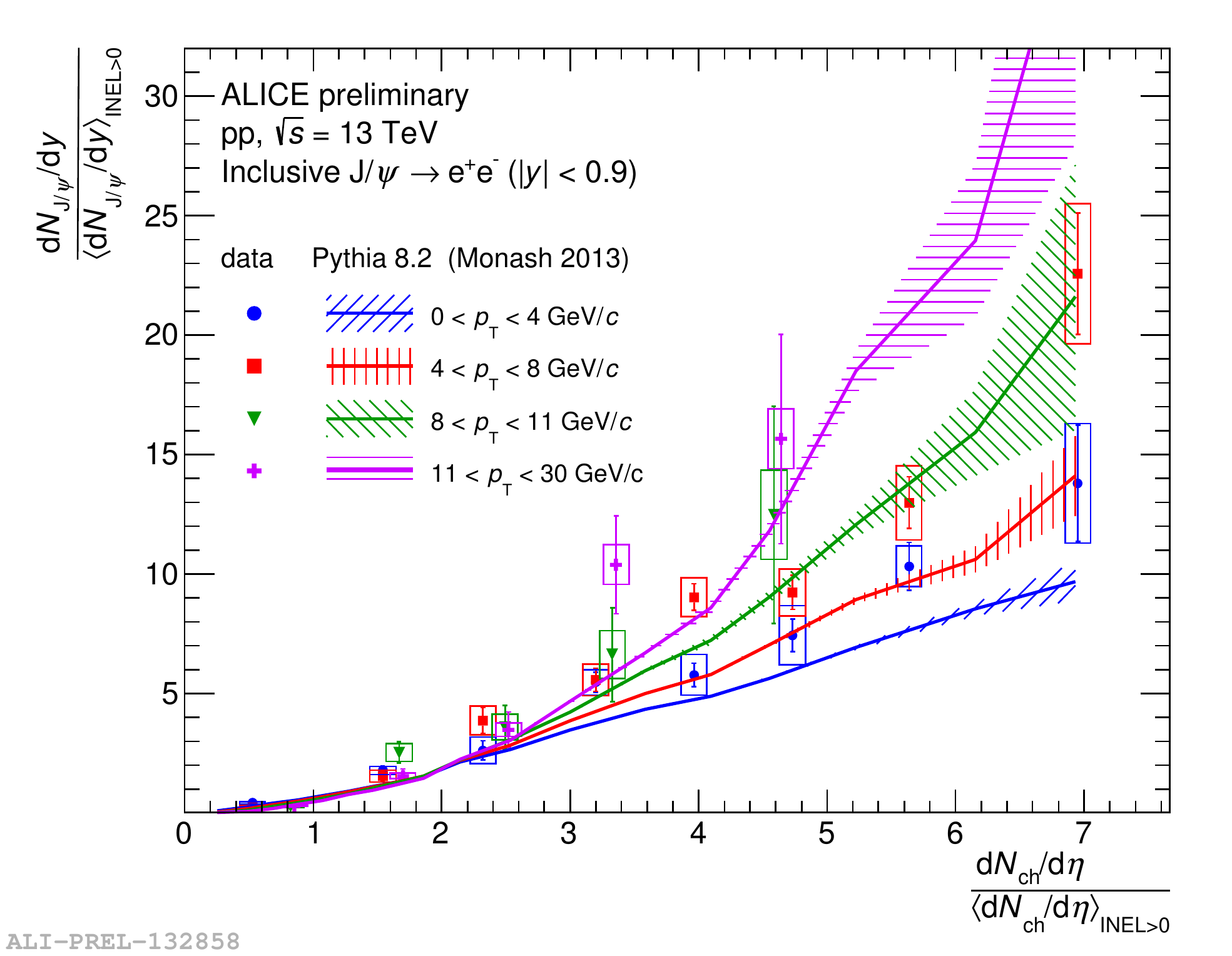}
\caption{\label{fig:result2} Multiplicity dependence of inclusive J/$\psi$ production at mid-rapidity in pp collisions at $\sqrt{s}$ = 13 TeV in different $p_{\rm T}$ intervals. The results are compared with PYTHIA 8  \cite{Sjostrand:2007gs}.}
\end{figure}

Figure \ref{fig:result2} shows the self-normalized inclusive J/$\psi$ as a function of self-normalized charged-particle multiplicity in pp collisions at $\sqrt{s}$ = 13 TeV for four different transverse momentum intervals. The two highest $p_{\rm T}$ intervals were obtained using the EMCal-triggered events and the two lowest $p_{\rm T}$ intervals were obtained using the ALICE high-multiplicity trigger. 
The increase of J/$\psi$ yield with respect to the charged-particle multiplicity is stronger than linear for all measured  $p_{\rm T}$ intervals and it is $p_{\rm T}$ dependent: a stronger rise is observed for higher $p_{\rm T}$ intervals.
The results are compared with PYTHIA 8 \cite{Sjostrand:2007gs} which reproduces qualitatively the increase and also the $p_{\rm T}$ dependence of our results. Since PYTHIA 8 includes MPI, this hints at an important role of multiple interactions at the parton level for charm production in hadron--hadron collisions.

\begin{figure}[h!]
\includegraphics[width=0.5\textwidth]{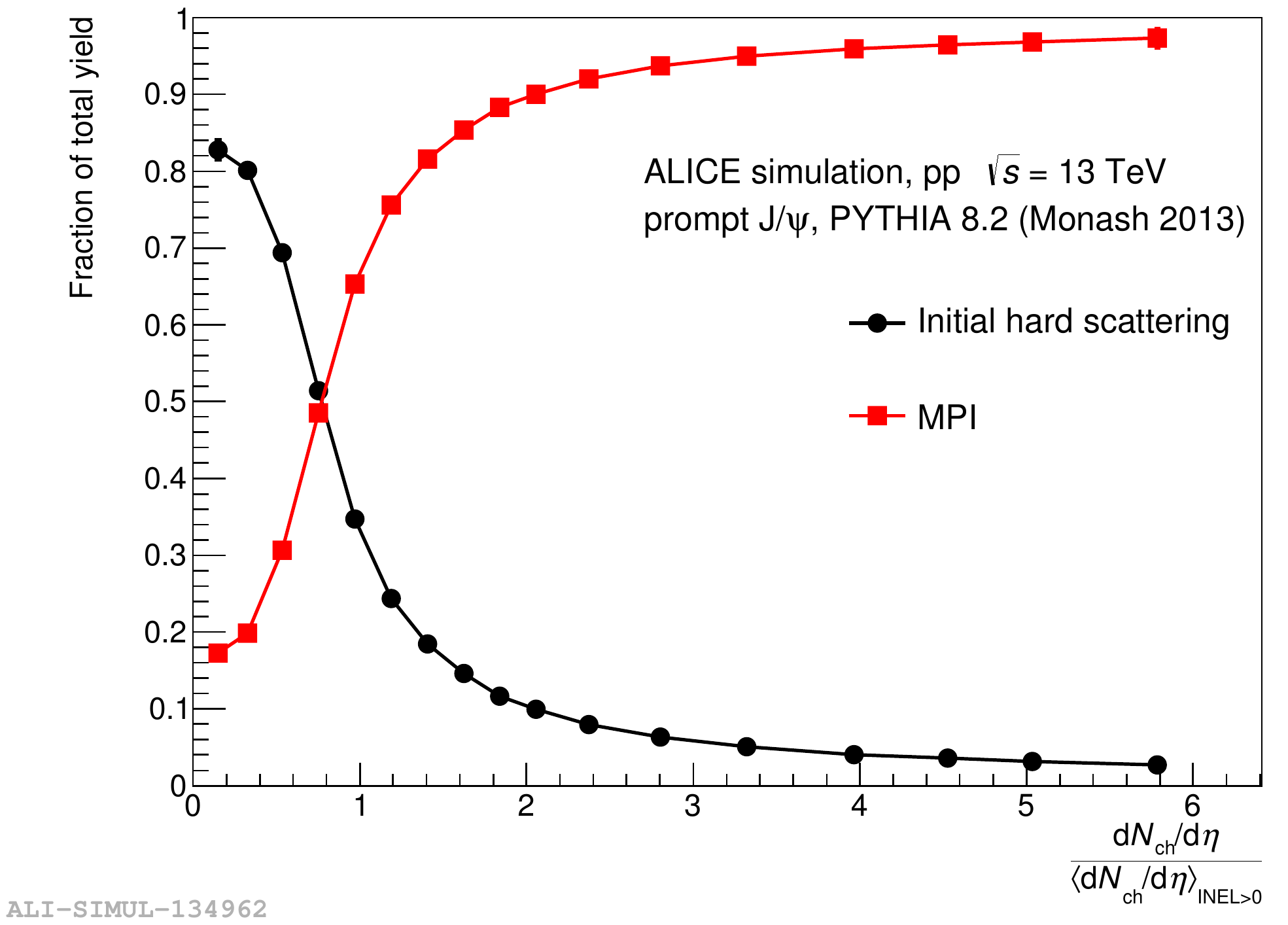}
\caption{\label{fig:sim} Contribution to total J/$\psi$ yield from initial scattering and MPI versus multiplicity from PYTHIA 8 in pp collisions at $\sqrt{s}$ = 13 TeV.}
\end{figure}

 PYTHIA 8 simulations predict that at high multiplicities most of the J/$\psi$ originate from MPI while at low multiplicities the dominat process is the initial hard scattering, as it is illustrated in Fig. \ref{fig:sim}.

A recent development from Ma et al.~\cite{Ma:2018bax}, based on the Color Glass Condensate (CGC) effective field theory to compute short distance charmonium cross-sections can describe  the $p_{\rm T}$ integrated results.
 This model describes the J/$\psi$ hadronization employing Nonrelativistic QCD (NRQCD) and an Improved Color Evaporation model.
In this model, with increasing event activity the J/$\psi$ production is dominated by the $^{3}\rm{S}_{1}^{[8]}$ state, which is consistent with an interpretation of the dominance of hard gluon fragmentation in J/$\psi$ hadronization. 

\section{\label{sec:level1}Conclusions}

The production of J/$\psi$ mesons as a function of event multiplicity in pp collisions at $\sqrt{s}$ = 13 TeV measured by ALICE has been presented. The results show an increase of the J/$\psi$ yield stronger than linear with respect to the charged-particle multiplicity and there is a hint of a higher increase for higher $p_{\rm T}$ intervals. The rise seen in the $p_{\rm T}$ integrated results are described qualitatively by theoretical models that include MPI and color reconnection (PYTHIA 8 \cite{Sjostrand:2007gs}), a hydrodynamical expansion of the system and MPI (EPOS 3 \cite{Drescher:2000ha}), the percolation model (Ferreiro et al. \cite{Ferreiro:2012fb}), higher Fock states (Kopeliovich et al. \cite{Kopeliovich:2013yfa}) and CGC $+$ NRQCD (Ma et al. \cite{Ma:2018bax}). Quantitatively, CGC $+$ NRQCD and EPOS3 can describe the results best, which can lead to different interpretations: the increase seen in data is either an effect of the dominance of hard gluon fragmentation or an effect of MPI in a system expanding hydrodynamically.
The $p_{\rm T}$ dependence of the J/$\psi$ production is qualitatively described by PYTHIA 8, which can be a hint of MPI affecting the production of J/$\psi$ in high multiplicity pp events. It will be interesting to see the above mentioned theoretical approaches differentially in $p_{\rm T}$.


\bibliography{Jpsi}

\end{document}